\documentclass[letterpaper]{article} 
\usepackage[preprint]{aaai2027}  
\usepackage[hyphens]{url}  
\usepackage{graphicx} 
\usepackage{natbib}  
\usepackage{caption} 
\usepackage{algorithm}

\usepackage{newfloat}
\usepackage{listings}
\DeclareCaptionStyle{ruled}{labelfont=normalfont,labelsep=colon,strut=off} 
\floatstyle{ruled}
\newfloat{listing}{tb}{lst}{}
\floatname{listing}{Listing}

\usepackage{booktabs}

\usepackage{makecell}
\usepackage{textcomp}
\usepackage{verbatim}
\usepackage{cite}
\usepackage{xcolor}         
\usepackage{amsmath}
\usepackage{algpseudocode}
\usepackage{tabularx}
\usepackage{multirow}
\usepackage{utfsym}
\usepackage{amsfonts} 
\usepackage[table]{xcolor}

\def\a{\textcolor{black}}

\newcommand{\methodname}{EOPA}

\title{Preference-Driven Online Adaptation for Personalized Interaction Initiation \\in Proactive AI Assistants}

\affiliations{
    \textsuperscript{\rm 1}Association for the Advancement of Artificial Intelligence\\

    1101 Pennsylvania Ave, NW Suite 300\\
    Washington, DC 20004 USA\\
    proceedings-questions@aaai.org
}

\author {
    Yufeng Wang\textsuperscript{\rm 1,\rm 2}\equalcontrib,
    Wei Zhang\textsuperscript{\rm 1}\equalcontrib,
    Zhiquan Wen\textsuperscript{\rm 1},
    Jinwu Hu\textsuperscript{\rm 1},
    Linhui Xiao\textsuperscript{\rm 2},
    Tianlu Pan\textsuperscript{\rm 2}, \\
    Qingfang Zheng\textsuperscript{\rm 2}\corresponding,
    Mingkui Tan\textsuperscript{\rm 1}\corresponding
}
\affiliations {
    \textsuperscript{\rm 1}South China University of Technology\\
    \textsuperscript{\rm 2}Pengcheng Laboratory\\
    \{yufeng6568, zw2177738821, sewenzhiquan, fhujinwu\}@gmail.com,
    \{xiaolh, pantl, zhengqf01\}@pcl.ac.cn, mingkuitan@scut.edu.cn
    \vspace{-4pt}
}

\begin{document}

\maketitle

\begin{abstract}
AI assistants are typically reactive, relying on users to initiate interactions. Proactive assistants go beyond this paradigm by autonomously initiating interactions based on users' activity contexts. However, appropriate interaction timing is user-specific and difficult to determine in advance, while online feedback offers valuable signals for personalization. Direct feedback-driven adaptation is therefore appealing, but remains challenging due to sparse interaction-worthy moments scattered across fine-grained user states. To address the issues, we propose Evidence-driven Online Preference Adaptation (\methodname), which grounds a user's interaction-timing preferences in measurable contextual evidence through two evidence carriers: temporal preference anchors and evidence-bearing activity prototypes. At each polling step, \methodname~derives temporal and activity evidence from the carriers through user-prior-smoothed evidence estimation and uncertainty-guided evidence scaling, and adaptively fuses the evidence for interaction-or-silence decisions. When interaction is selected, an LLM uses high-quality historical responses as demonstrations to generate a context-aware response that better reflects user preferences. \methodname~updates its evidence carriers and decision parameters from received online feedback without LLM-based reasoning or retraining. Extensive experiments on a ProPerSim-based benchmark show that \methodname~improves the interaction-timing F1 score by 19.80 points over the strongest baseline in our experiments, substantially reduces inference latency for both silence and interaction steps, and lowers the average daily adaptation time from 11.41 to 0.39 seconds.
\end{abstract}

\section{Introduction}
\label{sec:introduction}

\begin{figure}
    \centering
    \includegraphics[width=0.95\linewidth]{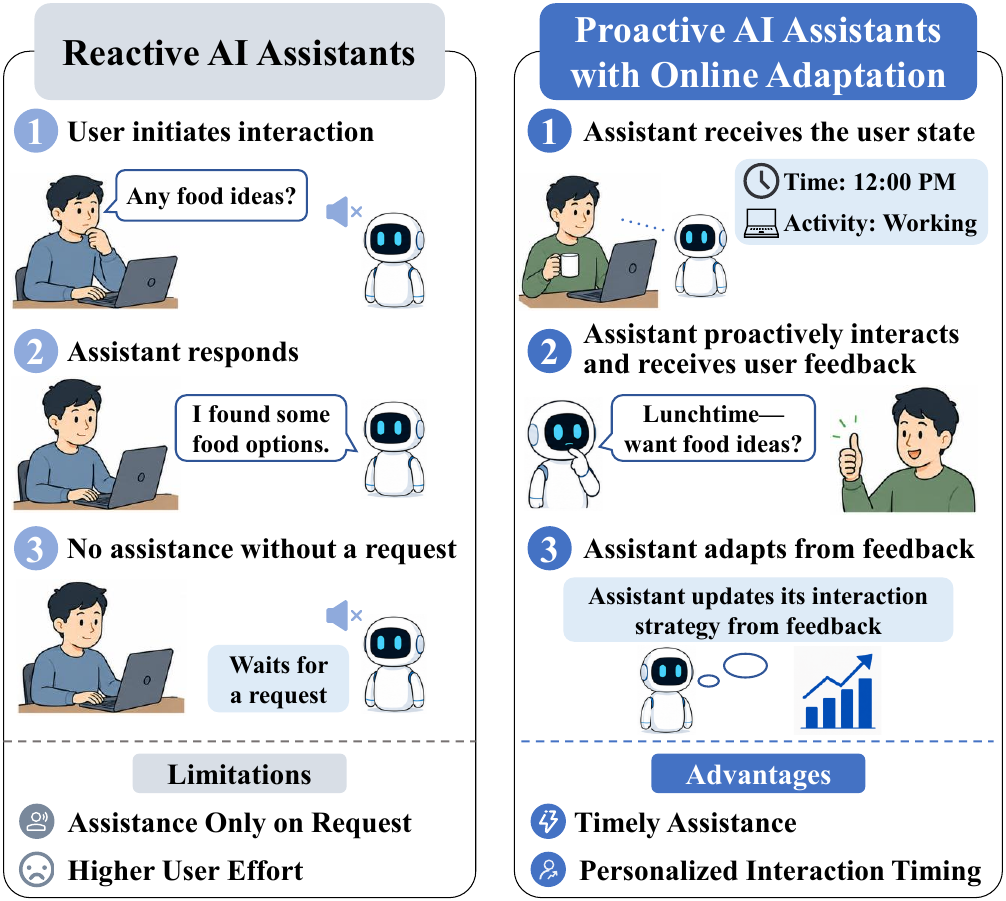}
    \caption{Comparison between reactive AI assistants and proactive AI assistants with online adaptation.}
    \label{fig:teaser}
    \vspace{-10pt}
\end{figure}

AI assistants provide information and suggestions to support users in various tasks~\cite{schobel2024charting}. However, most remain reactive, responding only after explicit user requests. Proactive assistants go beyond this reactive paradigm by autonomously initiating interactions at appropriate moments based on users’ activity contexts~\cite{luproactive}, as shown in Fig.~\ref{fig:teaser}. They can be deployed on embodied robots, smart glasses, in-vehicle systems, and other context-aware devices to provide timely support in applications such as virtual assistance~\cite{shrivastava2025natural,csigdigital}, healthcare~\cite{tu2025towards,habicht2024closing}, and accessibility support~\cite{algamdi2026behaviour}.
However, interaction-timing preferences are inherently personalized, as the same situation may warrant assistance for one user but be unnecessary for another.
Such preferences are difficult to determine in advance, but can be progressively revealed through user feedback from ongoing interactions. This raises a question: \textit{how can an assistant leverage such feedback for online adaptation to better align interaction timing with personalized user needs?}

This adaptation is difficult because \textit{interaction-worthy moments are sparse and scattered across fine-grained user states} (\textit{i.e.,} user activity and current time). 
Specifically, a proactive assistant continuously polls the user's state over time, but only a small fraction of the observed states may present useful opportunities for interaction, such as offering a relevant suggestion or timely reminder. \a{Moreover, similar states may require different interaction decisions, making their fine-grained interaction needs difficult to distinguish.}

To alleviate the above issues, recent approaches mainly follow two paradigms. Parameter-learning methods fine-tune the Large Language Models (LLMs) from contextual or interaction data~\cite{luproactive,yangcontextagent,kim2026propersim}.  However, sparse interaction-worthy moments are dispersed across diverse contexts, providing insufficient information for learning fine-grained interaction-timing preferences.
LLM-based self-improvement methods instead summarize past trajectories and feedback into natural-language reflections, rules, or configurations~\cite{zhao2024expel,shinn2023reflexion,pei2025scope,he2025evotest}. 
\a{Although they avoid gradient-based updates, such natural-language experience often fails to precisely specify its applicability boundaries, risking overgeneralization or overly narrow reuse.}

In this work, we raise a question: \emph{How can an assistant turn user feedback into reusable experience that preserves fine-grained differences in interaction needs among similar user states?}
Our key insight is to \a{ground a user's interaction-timing preferences in measurable contextual evidence. The assistant accumulates interaction and silence statistics under different temporal and activity-semantic conditions, and converts the statistics relevant to the current state into contextual evidence. This evidence quantifies the appropriateness of proactive interaction for the current user state and helps distinguish similar states with different interaction needs.}

Based on this insight, we propose \textbf{E}vidence-driven \textbf{O}nline \textbf{P}reference \textbf{A}daptation (\methodname) for personalized interaction initiation.
\methodname~organizes online feedback into two complementary evidence carriers: temporal preference anchors for time-conditioned interaction patterns and evidence-bearing activity prototypes for activity-semantic patterns. \methodname~then transforms the carrier statistics into temporal and activity evidence through user-prior-smoothed evidence estimation and uncertainty-guided evidence scaling, which stabilize estimates under limited local observations. The two evidence sources are adaptively fused and compared with an online-calibrated threshold for interaction-or-silence decisions. Response generation is decoupled from timing decisions: only when interaction is selected does an LLM generate a context-aware response using high-quality historical responses as in-context demonstrations.
After receiving user feedback, \methodname~updates the evidence carriers and recalibrates the fusion weight and decision threshold, enabling online adaptation of personalized interaction. Our contributions are summarized as follows:

\begin{itemize}

\item We propose \methodname, which grounds a user's interaction-timing preferences in measurable contextual evidence. It organizes online feedback through two evidence carriers, temporal preference anchors and evidence-bearing activity prototypes, to preserve reusable evidence across fine-grained temporal and activity-semantic conditions.

\item We formulate temporal and activity evidence through user-prior-smoothed evidence estimation and uncertainty-guided evidence scaling, which stabilize estimates under limited local observations. The evidence carriers and decision parameters are further adapted from accumulated feedback without LLM-based reasoning or retraining, enabling efficient online personalization.

\item  Extensive experiments on a ProPerSim-based benchmark show that \methodname~improves the interaction-timing F1 score by 19.80 points over the Reflexion baseline and substantially reduces inference latency and adaptation time.

\end{itemize}

\section{Related Work}

\subsection{Proactive Assistants}

\textbf{Rule-Based Methods.}
Early systems determine when to deliver information or initiate assistance using manually specified conditions and triggering heuristics~\cite{dey2000cybreminder,siewiorek2003sensay,ho2005using,dorneich2012considering}. These rules typically consider factors such as time, location, activities, user workload, and message priority. However, these methods are inflexible and limited to manually predefined conditions.

\textbf{Parameter-Learning Methods.}
Recent methods construct contextual or interaction data and optimize models to predict users' needs for proactive assistance~\cite{yangcontextagent,luproactive,kim2026propersim,yang2025proagent}. ProactiveAgent and ContextAgent construct context-aware training data and fine-tune LLMs to predict proactive actions or tool calls~\cite{luproactive,yangcontextagent}, while ProPerSim provides persona-conditioned user simulation and feedback for continually improving personalized proactive assistants~\cite{kim2026propersim}. However, sparse interaction-worthy moments provide limited supervision for learning fine-grained timing preferences. Repeated weight updates also introduce substantial adaptation costs.

\textbf{LLM-Based Self-Improvement Methods.}
LLM-based self-improvement methods avoid weight updates by summarizing past trajectories and feedback into natural-language reflections, guidelines, or configurations~\cite{zhao2024expel,shinn2023reflexion,pei2025scope,he2025evotest,he2026evoclinician}. For example, ExpeL~\cite{zhao2024expel} and Reflexion~\cite{shinn2023reflexion} accumulate textual experience from previous trajectories. SCOPE~\cite{pei2025scope} and EvoTest~\cite{he2025evotest} update prompts or agent configurations during test-time interaction. However, natural-language experience usually expresses applicability conditions only implicitly, making it difficult to distinguish fine-grained differences in interaction needs among similar user states.

\begin{figure*}
    \centering
    \includegraphics[width=0.95\linewidth]{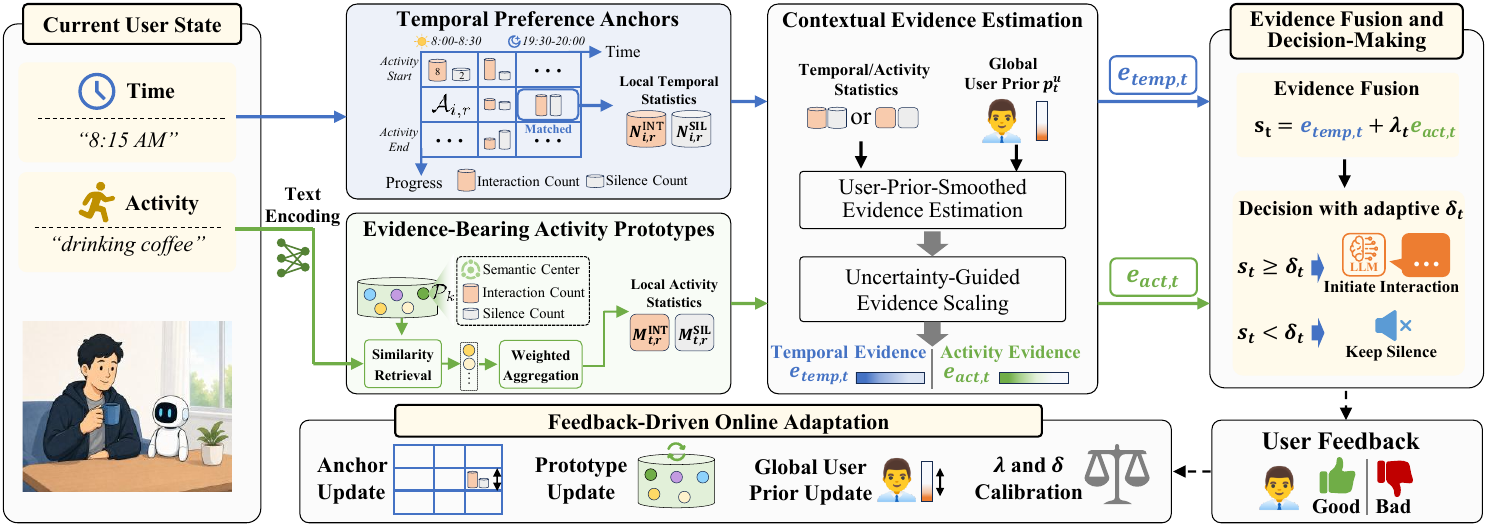}
    \caption{\methodname~grounds a user's interaction-timing preferences in measurable contextual evidence from two carriers: temporal preference anchors and evidence-bearing activity prototypes. Given the current time and user activity, it retrieves the relevant carrier statistics and estimates temporal and activity evidence through user-prior-smoothed evidence estimation and uncertainty-guided scaling, and adaptively fuses the evidence for interaction decisions. When interaction is selected, an LLM generates a context-aware response. \methodname~then uses user feedback to update evidence carriers and decision parameters online.}
\label{fig:method}
\vspace{-10pt}
\end{figure*}

\subsection{Context-Aware Preference Modeling}

Context-aware preference modeling estimates user preferences by considering relevant contextual conditions~\cite{pitis2024improving}. Existing methods model context-specific preferences based on explicitly specified contexts~\cite{pitis2024improving}, user-specified values~\cite{lee2024aligning}, or inferred user-specific latent preferences~\cite{poddar2024personalizing,gao2024aligning}. Other studies capture temporal patterns or retrieve context-relevant historical behaviors~\cite{chiang2022adaptive,yu2023continuous,lin2024rella}. These studies demonstrate that user preferences should be modeled in relation to context rather than
solely through a global user profile. Nevertheless, most focus on determining what to recommend or how to respond, with limited attention to interaction timing.

\section{Task Formulation and Motivation}
\label{sec:task_definition}

\textbf{Problem Formulation.}
We consider a proactive conversational AI assistant that accompanies a user, continuously receives the user's activity state, and autonomously initiates interactions without explicit user request.
In practice, activities can be represented as textual descriptions by streaming video captioning~\cite{zhou2024streaming,li2026challenges}.
Building on such perceptual outputs, we study how the assistant makes interaction-or-silence decisions over time.
At each polling step $t$, the assistant observes the current user state
\begin{equation}
x_t = (\tau_t, a_t),
\end{equation}
where $\tau_t$ denotes the current time and $a_t$ denotes the textual description of the user's current activity.
It then makes an interaction-or-silence decision
\begin{equation}
\hat{y}_t \in \{\mathrm{INT},\mathrm{SIL}\},
\end{equation}
where $\mathrm{INT}$ and $\mathrm{SIL}$ denote initiating an interaction and remaining silent, respectively.
If $\hat{y}_t=\textsc{INT}$, the assistant generates a context-aware response; otherwise, it remains silent.
After each decision, the assistant receives user feedback, from which a timing label
$y_t\in\{\mathrm{INT},\mathrm{SIL}\}$ is derived.
Interaction steps additionally receive response-quality feedback. The assistant uses these signals to refine its personalized interaction strategy.

\textbf{Motivation.}
Existing methods either learn from sparse interaction supervision or summarize feedback as natural-language experience with implicit applicability conditions, making fine-grained interaction needs among similar states difficult to distinguish. Our motivation is to transform historical feedback into updatable and measurable contextual evidence. Rather than compressing feedback into textual experience, the assistant stores interaction and silence statistics under specific temporal and activity-semantic conditions. When a new state arrives, it retrieves the matched temporal anchor and semantically relevant activity prototypes, and uses their condition-specific statistics to estimate how strongly the current state supports proactive interaction. As a result, states that appear similar but differ in contextual conditions can draw on different evidence, allowing historical feedback to be reused without losing fine-grained differences among different interaction needs.

\section{Method}
\label{sec:method}
\a{We propose \methodname, an evidence-driven framework for personalized proactive interaction. The key idea is to ground a user's interaction-timing preferences in measurable contextual evidence, thereby preserving different interaction needs among similar user states. As shown in Fig.~\ref{fig:method}, temporal preference anchors and evidence-bearing activity prototypes capture complementary temporal and activity-semantic patterns. Their statistics are transformed into temporal and activity evidence through user-prior-smoothed evidence estimation and uncertainty-guided scaling, and the resulting evidence is adaptively fused for interaction decisions. When interaction is selected, an LLM generates a context-aware response using historical responses as in-context demonstrations. Online adaptation updates the evidence carriers and decision parameters from feedback without LLM-based reasoning or retraining. The pseudocode is provided in Appendix~A.}

\subsection{Temporal Preference Anchors}
\label{sec:temporal_preference_anchors}

Interaction appropriateness often varies across different periods of the day.
Within the same period, it may also change as an activity progresses; for example, assistance may be useful at the beginning of an activity but disturbing after the user becomes engaged.
\methodname~therefore maintains temporal preference anchors indexed by both time interval and activity-progress stage.

\textbf{Anchor Construction.}
We divide each day into $B$ fixed-width time intervals,
$\mathcal{I}=\{i_1,i_2,\ldots,i_B\}$,
and define the progress-index set
$\mathcal{R}=\{1,2,\ldots,R\}$.
At polling step $t$, the current time $\tau_t$ determines an interval
$i\in\mathcal{I}$.
The progress index $r\in\mathcal{R}$ denotes the number of consecutive polling steps elapsed within the current activity segment.
For each $(i,r)\in\mathcal{I}\times\mathcal{R}$, the temporal preference anchor is defined as
\begin{equation}
\mathcal{A}_{i,r}
=
\left(
N_{i,r}^{\mathrm{INT}},
N_{i,r}^{\mathrm{SIL}}
\right),
\end{equation}
where $N_{i,r}^{\mathrm{INT}}$ and $N_{i,r}^{\mathrm{SIL}}$ denote the accumulated counts of observations labeled as interaction-appropriate and silence-preferred at time interval $i$ and progress index $r$, respectively.

\textbf{User-Prior-Smoothed Evidence Estimation.} 
For the matched anchor $\mathcal{A}_{i,r}$, $N_{i,r}^{\mathrm{INT}}$ and $N_{i,r}^{\mathrm{SIL}}$ record how often interaction and silence are appropriate under the corresponding time-progress condition. Their empirical ratio provides a direct estimate of interaction appropriateness, but can be unreliable when the anchor contains only a few observations. \a{Because interaction timing involves a binary interaction-or-silence decision, we use a Beta--Binomial formulation~\cite{gelman2013bayesian} to obtain a user-prior-smoothed estimate of interaction appropriateness for the matched anchor.} Before step $t$, this prior is computed using the Jeffreys-smoothed empirical rate~\cite{jeffreys1946invariant}:
\begin{equation}
p_t^{u}
=
\frac{
C_{t-1}^{\mathrm{INT}}+\frac{1}{2}
}{
C_{t-1}^{\mathrm{INT}}
+
C_{t-1}^{\mathrm{SIL}}
+
1
},
\label{eq:user_interaction_prior}
\end{equation}
where $C_{t-1}^{\mathrm{INT}}$ and $C_{t-1}^{\mathrm{SIL}}$ are the accumulated interaction-appropriate and silence-preferred counts, respectively.
We use $p_t^u$ as the prior mean and estimate the interaction-appropriateness probability for the matched anchor as
\begin{equation}
p_{\mathrm{temp},t}
=
\frac{
N_{i,r}^{\mathrm{INT}}
+
\kappa p_t^{u}
}{
N_{i,r}^{\mathrm{INT}}
+
N_{i,r}^{\mathrm{SIL}}
+
\kappa
},
\label{eq:temporal_interaction_estimate}
\end{equation}
where $\kappa>0$ controls the strength of prior smoothing. The user-level prior stabilizes the estimate when the matched anchor contains only a few observations, while anchor-specific counts increasingly dominate as local feedback accumulates. \a{The Beta--Binomial derivation is provided in Appendix~B.}

\textbf{Uncertainty-Guided Evidence Scaling.}
The posterior mean $p_{\mathrm{temp},t}$ estimates interaction appropriateness under the matched time-progress condition, but does not indicate how strongly the estimate is supported by local feedback. Anchors with similar posterior means may contain substantially different numbers of observations, making sparsely supported estimates more uncertain.
We therefore quantify this uncertainty using the corresponding Beta posterior variance~\cite{gelman2013bayesian}:
\begin{equation}
\sigma_{\mathrm{temp},t}^{2}
=
\frac{
p_{\mathrm{temp},t}
\left(1-p_{\mathrm{temp},t}\right)
}{
N_{i,r}^{\mathrm{INT}}+N_{i,r}^{\mathrm{SIL}}+\kappa+1
}.
\label{eq:temporal_preference_uncertainty}
\end{equation}
As feedback accumulates at the matched anchor, the variance decreases.
We then scale the posterior mean as follows:
\begin{equation}
e_{\mathrm{temp},t}=\frac{
p_{\mathrm{temp},t}
}{
\sqrt{\sigma_{\mathrm{temp},t}^{2}+\epsilon}
},
\label{eq:temporal_preference_evidence}
\end{equation}
where $\epsilon=10^{-12}$ ensures numerical stability.
The resulting evidence score tends to be larger when the temporal condition favors interaction and receives stronger support from accumulated feedback, allowing better-supported anchors to contribute more strongly.

\textbf{Anchor Update.}
At every step, EOPA derives timing label
$y_t$ from user feedback and updates the matched anchor as
\begin{equation}
N_{i,r}^{d}
\leftarrow
N_{i,r}^{d}
+
\mathbb{I}\!\left(y_t=d\right),
\quad
d\in\{\mathrm{INT},\mathrm{SIL}\},
\label{eq:temporal_anchor_update}
\end{equation}
where $\mathbb{I}(\cdot)$ is the indicator function, equal to $1$ when the condition holds and $0$ otherwise.
The user-level counts are updated in the same manner for estimating the interaction prior at subsequent polling steps.
Each temporal preference anchor thereby accumulates feedback under its matched time-progress condition for future evidence estimation.

\subsection{Evidence-Bearing Activity Prototypes}
\label{sec:semantic_activity_prototypes}

Temporal preference anchors distinguish interaction patterns across time and activity progress, but cannot separate different activities occurring under the same temporal condition.
Since activity descriptions are open-ended, \methodname~uses semantic prototypes that can be expanded online as new activities are observed~\cite{wei2023online}.
Unlike conventional prototypes that mainly serve as semantic centroids, each prototype in \methodname~is evidence-bearing: it stores both a semantic activity center and progress-conditioned interaction/silence statistics.
The semantic center supports relevance matching, while the stored statistics support reusable evidence estimation for
interaction decisions under related activity states.

\textbf{Prototype Representation.}
At step $t$, a frozen text encoder $\phi(\cdot)$ maps the activity description $a_t$ to a semantic representation:
$\mathbf{z}_t=\phi(a_t)$.
It is used to retrieve semantically related prototypes.
The $k$-th  prototype is defined as
\begin{equation}
\mathcal{P}_k
=
\left(
\mathbf{c}_k,
\left\{
\left(
H_{k,r}^{\mathrm{INT}},
H_{k,r}^{\mathrm{SIL}}
\right)
\right\}_{r\in\mathcal{R}}
\right),
\label{eq:activity_prototype}
\end{equation}
where $\mathbf{c}_k$ is the semantic center, and each prototype
maintains $H_{k,r}^{\mathrm{INT}}$ and
$H_{k,r}^{\mathrm{SIL}}$ for every $r\in\mathcal{R}$,
recording its interaction-appropriate and silence-preferred counts, respectively.

\textbf{Relevance-Weighted Prototype Aggregation.}
Let
$q_{t,k}=\operatorname{sim}\left(\mathbf{z}_t,\mathbf{c}_k\right)$
denote the semantic similarity between the current activity and prototype $k$.
We select the $K$ most similar prototypes and denote their indices by $\mathcal{S}_t$.
Because semantically related activities may still have different interaction needs, each retrieved prototype is weighted according to its relevance to the current activity:
\begin{equation}
w_{t,j}
=
\exp\left(
\frac{q_{t,j}-q_t^{\max}}{\gamma}
\right)
\left[\max\left(0,q_t^{\max}\right)\right]^2,
\label{eq:semantic_applicability_weight}
\end{equation}
where $q_t^{\max}$ is the largest similarity score among the retrieved prototypes and $\gamma>0$ is a temperature coefficient.
The exponential term emphasizes more similar prototypes, while the coverage term suppresses evidence transfer when even the closest prototype is weakly related to the current activity.
The activity-conditioned statistics are then aggregated as
\begin{equation}
M_{t,r}^{\mathrm{INT}}
=
\sum_{j\in\mathcal{S}_t}
w_{t,j}H_{j,r}^{\mathrm{INT}},
\quad
M_{t,r}^{\mathrm{SIL}}
=
\sum_{j\in\mathcal{S}_t}
w_{t,j}H_{j,r}^{\mathrm{SIL}}.
\label{eq:activity_statistics}
\end{equation}
These weighted statistics summarize interaction and silence feedback from activities relevant to the current state.

\textbf{Activity Evidence Estimation.}
Following the temporal branch, we estimate interaction appropriateness from the aggregated activity statistics using the global user prior:
\begin{equation}
p_{\mathrm{act},t}
=
\frac{
M_{t,r}^{\mathrm{INT}}
+
\kappa p_t^u
}{
M_{t,r}^{\mathrm{INT}}
+
M_{t,r}^{\mathrm{SIL}}
+
\kappa
}.
\label{eq:activity_interaction_estimate}
\end{equation}
The corresponding posterior variance is
\begin{equation}
\sigma_{\mathrm{act},t}^{2}
=
\frac{
p_{\mathrm{act},t}
\left(1-p_{\mathrm{act},t}\right)
}{
M_{t,r}^{\mathrm{INT}}
+
M_{t,r}^{\mathrm{SIL}}
+
\kappa
+
1
}.
\label{eq:activity_uncertainty}
\end{equation}
The activity evidence is then obtained as
\begin{equation}
e_{\mathrm{act},t}
=
\frac{
p_{\mathrm{act},t}
}{
\sqrt{\sigma_{\mathrm{act},t}^{2}+\epsilon}
}.
\label{eq:activity_evidence}
\end{equation}
A larger $e_{\mathrm{act},t}$ indicates stronger activity-semantic support for interaction initiation.

\textbf{Prototype Update.}
After receiving the feedback-derived timing label, \methodname~selects the most similar existing prototype if $q_t^{\max}\geq\rho$; otherwise, it creates a new prototype, where $\rho$ is the similarity threshold.
Let $\mathcal{P}_{\tilde{k}}$ denote the prototype for update.
Its semantic center is updated by incremental averaging if it already exists, or initialized as $\mathbf{z}_t$ if newly created, while its statistics are updated as
\begin{equation}
H_{\tilde{k},r}^{d}
\leftarrow
H_{\tilde{k},r}^{d}
+
\mathbb{I}(y_t=d),
\quad
d\in\{\mathrm{INT},\mathrm{SIL}\}.
\label{eq:prototype_feedback_update}
\end{equation}
Thus, newly received feedback is incorporated into reusable activity-semantic evidence for subsequent decisions.

\subsection{Evidence Fusion and Adaptive Decisions}
\label{sec:evidence_fusion_update}

The temporal and activity branches provide complementary evidence for proactive interaction.
Their outputs are fused as
\begin{equation}
s_t
=
e_{\mathrm{temp},t}
+
\lambda_t e_{\mathrm{act},t},
\label{eq:fusion_score}
\end{equation}
where $\lambda_t$ controls the contribution of activity evidence.
The interaction decision is then made using threshold $\delta_t$:
\begin{equation}
\hat{y}_t=
\mathrm{INT}\ \text{if }s_t\geq\delta_t,
\ \mathrm{SIL}\ \text{otherwise}.
\label{eq:final_decision}
\end{equation}

Since users may differ in how strongly temporal and activity evidence indicate their interaction needs and in their receptiveness to proactive assistance, \methodname~adapts $\lambda_t$ and $\delta_t$ online.
It maintains a recent history of temporal evidence, activity evidence, and feedback labels, periodically replays decisions under candidate parameter pairs, and adjusts the parameters to improve the F1 score on the historical data.

\subsection{Response Generation}
When interaction is selected, an LLM generates a response based on the current activity and communication-style cues from recent interactions.
High-quality historical responses are selected as in-context demonstrations.
When no suitable examples are available, the LLM follows a concise, polite, and context-aware default style. By decoupling interaction timing decisions from response generation, \methodname~allows the LLM to focus solely on response quality using a contextual prompt, as detailed in Appendix~C.

\section{Experiments}
\subsection{Experimental Settings}

\textbf{Benchmark and Online Protocol.}
We build on ProPerSim~\cite{kim2026propersim}, which provides 32 diverse user personas whose daily activities are simulated using the Generative Agents framework~\cite{park2023generative}. Following its protocol, each user is simulated for 14 days, yielding 448 user-days and 160,632 polling steps at 2.5-minute intervals. At each step, the assistant accesses only the current state and historical interactions, decides whether to interact or remain silent, and then receives binary timing feedback, with additional response-quality feedback for interaction steps. Details of the benchmark are provided in Appendix~D.

\textbf{Timing Annotation.}
For reproducible evaluation, we construct fixed interaction-timing labels with Qwen3.5-397B-A17B~\cite{qwen3.5}, conditioned on the timing and frequency preferences provided by ProPerSim. The LLM first estimates hourly interaction frequency and then selects specific steps. The detailed annotation process is provided in Appendix~E. This procedure produces 1,814 interaction-appropriate moments, accounting for only 1.129\% of all polling steps. We further conduct a human evaluation to verify the quality of the simulated activities and timing annotations. The protocol and results are provided in Appendix~F.

\textbf{Metrics.}
Following ProPerSim, we evaluate Timing, Frequency, Personal Preference, and Communication \& Safety. For Timing, we report F1 score, Accuracy, Precision, and Recall by comparing interaction-or-silence predictions with the annotated timing labels. We also report F1@1 and F1@2, which count predictions within one and two polling steps of an annotated interaction moment as correct. Given the severe class imbalance, we treat the F1-based metrics as the primary metrics rather than Accuracy~\cite{he2009learning}. For Frequency, we compute the proportion of hours whose predicted interaction count matches the expected count. We use Qwen3.5-397B-A17B as the judge LLM to evaluate Personal Preference and Communication \& Safety. 
All metrics are computed per user over the 14-day trajectory and averaged across 32 users.
More details are in Appendix~G.

\textbf{Baselines.}
We compare \methodname~with ProPerAssistant~\cite{kim2026propersim}, EvoTest~\cite{he2025evotest}, SCOPE~\cite{pei2025scope}, and Reflexion~\cite{shinn2023reflexion}. All methods use the same polling interval, feedback protocol, evaluation prompts, and Qwen3-4B-Instruct-2507~\cite{yang2025qwen3} as the response-generation LLM.
In the main comparison, the LLM-based self-improvement baselines use Qwen3-4B-Instruct-2507, DeepSeek-V4-Flash~\cite{xu2026deepseek}, and
GLM-5~\cite{zeng2026glm} as adaptation LLMs, while further analyses use Qwen3-4B. More details are in Appendix~H.

\begin{table*}[t]
\centering
\scriptsize
\caption{Comparison results. Qwen3-4B means Qwen3-4B-Instruct-2507 and DeepSeek-V4 means DeepSeek-V4-Flash.}
\label{tab:main_results}
\setlength{\tabcolsep}{3.0pt}
\begin{tabular}{cccccccccccc}
\toprule
Category & Method & Adaptation LLM 
& F1 score $\uparrow$ & F1\symbol{64}1 $\uparrow$ & F1\symbol{64}2 $\uparrow$ 
& Recall $\uparrow$ & Precision $\uparrow$ & Accuracy $\uparrow$ & Freq. $\uparrow$ & Pref. $\uparrow$ & Comm.\&Safe. $\uparrow$\\
\midrule

\multirow{1}{*}{Parameter Learning}
& \makecell{ProPerAssistant\\\cite{kim2026propersim}} & None
& 2.57 & 4.15 & 4.98 & 28.82 & 1.36 & 75.35 & 9.84 & 45.12 & 87.22 \\

\midrule

\multirow{9}{*}{\makecell{LLM-based\\Self-Improvement}}

& \multirow{3}{*}{\makecell{EvoTest\\\cite{he2025evotest}}}
& Qwen3-4B
& 8.29 & 9.42 & 10.12 & 25.24 & 6.27 & 94.20 & 46.79 & 44.28 & 91.24 \\
& & DeepSeek-V4
& 8.73 & 9.85 & 10.56 & 29.88 & 5.87 & 93.55 & 42.34 & 47.78 & 92.64 \\
& & GLM-5
& 8.53 & 9.97 & 10.87 & 20.40 & 6.10 & 95.58 & 54.24 & 50.24 & 88.78 \\

\cmidrule(lr){2-12}

& \multirow{3}{*}{\makecell{SCOPE\\\cite{pei2025scope}}}
& Qwen3-4B
& 5.45 & 8.37 & 10.35 & 13.48 & 3.86 & 95.12 & 42.63 & 40.54 & 90.06 \\
& & DeepSeek-V4
& 5.62 & 8.29 & 9.66 & 9.79 & 4.43 & 96.88 & 56.35 & 41.09 & 88.47 \\
& & GLM-5
& 2.18 & 3.07 & 3.44 & 2.52 & 5.01 & 97.66 & 67.72 & 39.45 & 91.46 \\

\cmidrule(lr){2-12}

& \multirow{3}{*}{\makecell{Reflexion\\\cite{shinn2023reflexion}}}
& Qwen3-4B
& 10.44 & 11.69 & 12.93 & 25.17 & 6.82 & 95.42 & 43.77 & 41.41 & 70.20 \\
& & DeepSeek-V4
& 10.66 & 11.90 & 13.14 & 24.05 & 7.04 & 95.76 & 46.33 & 38.60 & 73.31 \\
& & GLM-5
& 10.07 & 11.41 & 12.69 & 22.04 & 6.68 & 95.81 & 46.61 & 37.59 & 73.68 \\

\midrule
Evidence-Driven Adaptation
& \textbf{\methodname~(Ours)} & None
& \textbf{30.46} & \textbf{32.77} & \textbf{33.54}
& \textbf{41.10} & \textbf{24.89}
& \textbf{97.73}
& \textbf{71.08}
& \textbf{52.65}
& \textbf{92.71} \\

\bottomrule
\end{tabular}
\vspace{-5pt}
\end{table*}

\begin{table}[t]
\centering
\caption{Ablation studies of \methodname.}
\label{tab:incremental_ablation}
\scriptsize
\setlength{\tabcolsep}{2.8pt}
\begin{tabular}{ccccc|cccc}
\toprule
GUP & T-Anch. & T-ED & A-Branch & Ada.
& F1 $\uparrow$ & F1@1 $\uparrow$ & F1@2 $\uparrow$ & Accuracy $\uparrow$\\
\midrule

\usym{1F5F8} & -- & -- & -- & --
& 1.13 & 2.63 & 2.63 & 87.68 \\

\usym{1F5F8} & \usym{1F5F8} & -- & -- & --
& 17.57 & 19.15 & 19.80 & 92.44 \\

\usym{1F5F8} & \usym{1F5F8} & \usym{1F5F8} & -- & --
& 25.81 & 27.94 & 28.57 & \textbf{98.21} \\

\usym{1F5F8} & \usym{1F5F8} & \usym{1F5F8} & \usym{1F5F8} & --
& 28.95 & 31.50 & 32.39 & 97.62 \\

\usym{1F5F8} & \usym{1F5F8} & \usym{1F5F8} & \usym{1F5F8} & \usym{1F5F8}
& \textbf{30.46} & \textbf{32.77} & \textbf{33.54} & 97.73 \\

\bottomrule
\end{tabular}
\vspace{-5pt}
\end{table}

\textbf{Implementation Details.}
We divide each day into 30-min intervals and set $\kappa=2.5$ for both evidence branches. 
The activity branch uses a frozen \texttt{all-mpnet-base-v2} encoder~\cite{all_mpnet_base_v2}, with $K=12$, $\gamma=0.1$ and $\rho=0.96$. The timing-decision pipeline is deterministic, and Qwen3-4B-Instruct-2507 generates responses at temperature 0.2 upon interaction. More details and hyperparameter analysis are in Appendix~I. Per-user performance distributions and paired significance tests are in Appendix~J.

\subsection{Comparison Experiments}
Table~\ref{tab:main_results} shows that \methodname~outperforms the baselines in both interaction timing and response quality. For interaction timing, \methodname~achieves 30.46 F1, 32.77 F1@1, and 33.54 F1@2. Compared with the best-performing Reflexion using Qwen3-4B and DeepSeek-V4 as adaptation LLMs, \methodname~improves F1 by 20.02 and 19.80 points, respectively.
ProPerAssistant attains 28.82 Recall but only 1.36 Precision, indicating that parameter updates struggle to localize sparse interaction-worthy moments and cause excessive interactions. LLM-based self-improvement methods reach at most 10.66 F1 and 7.04 Precision, suggesting limitations in precisely specifying the applicability boundaries of natural-language experience across similar user states with different interaction needs. Similar results across different adaptation LLMs suggest that experience applicability, rather than LLM capability, is an important bottleneck. In contrast, \methodname~achieves 24.89 Precision and 41.10 Recall, better limiting unnecessary interactions while identifying more appropriate moments.

Beyond interaction timing, \methodname~achieves the best Personal Preference and Communication \& Safety scores of 52.65 and 92.71. By decoupling timing decisions from response generation, the LLM can focus solely on response quality, achieving strong results with only simple in-context demonstrations from historical responses.

\subsection{Ablation Study}

We conduct an incremental ablation study to evaluate the contribution of each component in \methodname. As in Table~\ref{tab:incremental_ablation}, we begin with a minimal baseline that makes decisions solely from the global user prior without evidence carriers, denoted as \textbf{GUP}. We then progressively introduce temporal preference anchors (\textbf{T-Anch.}), \a{temporal evidence derivation} (\textbf{T-ED}), the activity-evidence branch (\textbf{A-Branch}), and adaptive fusion and decisions (\textbf{Ada.}) to obtain the full \methodname.

\textbf{Effectiveness of Temporal Evidence.}
GUP achieves only 1.13 F1, indicating that global interaction propensity cannot capture fine-grained timing preferences. Temporal preference anchors raise F1 to 17.57 by organizing feedback under specific time-progress conditions, while user-prior-smoothed evidence estimation and uncertainty-guided evidence scaling further improve it to 25.81 by stabilizing sparse anchor-level estimates and accounting for their uncertainty.

\textbf{Effectiveness of Activity-Semantic Evidence.}
Adding the activity-evidence branch improves F1 from 25.81 to 28.95, showing that activity-semantic evidence provides additional information beyond temporal evidence. Although Accuracy slightly decreases from 98.21 to 97.62, the consistent gains in the primary F1-based metrics demonstrate the overall benefit of incorporating activity-semantic conditions.

\textbf{Effectiveness of Adaptive Fusion and Decisions.}
Recalibrating the activity-evidence weight and decision threshold improves F1 from 28.95 to 30.46. This gain shows that adapting both evidence fusion and the decision threshold to accumulated feedback further improves interaction decisions.

\subsection{Further Experimental Analysis}

\begin{figure}[t]
    \centering
    \includegraphics[width=0.9\linewidth]{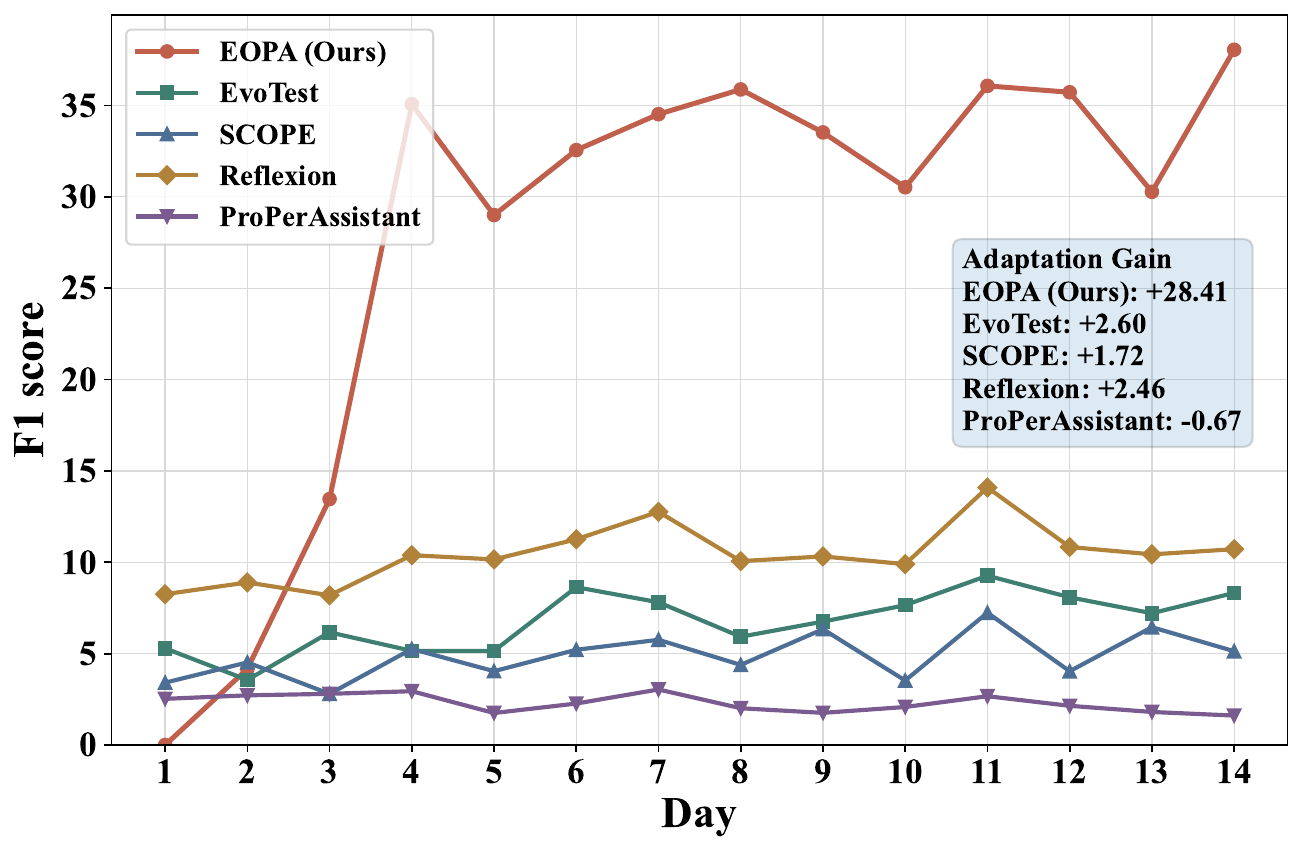}
    \caption{User-averaged daily F1 over the 14-day online stream. Adaptation gain is defined as the change in mean F1 from Days 1--3 to Days 8--14.}
    \label{fig:online_adaptation_curve}
    \vspace{-5pt}
\end{figure}

\textbf{Online Adaptation Dynamics.}
Fig.~\ref{fig:online_adaptation_curve} reports the daily F1 score averaged across 32 users over the 14-day interaction stream.
\methodname~rises sharply during the first four days and thereafter maintains a high-level performance. Its average F1 increases by 28.41 points from Days 1--3 to Days 8--14, whereas all baselines change by at most 2.60 points.
The baselines show limited gains: ProPerAssistant updates model parameters from sparse interaction supervision, which provides limited information for learning fine-grained interaction-timing preferences. Self-improvement methods encode feedback as natural-language experience with imprecise applicability boundaries. In contrast, \methodname~organizes accumulated feedback into fine-grained contextual evidence, and its sustained gains suggest that this evidence becomes increasingly useful as feedback accumulates.

\begin{figure}[t]
    \centering
    \includegraphics[width=0.9\linewidth]{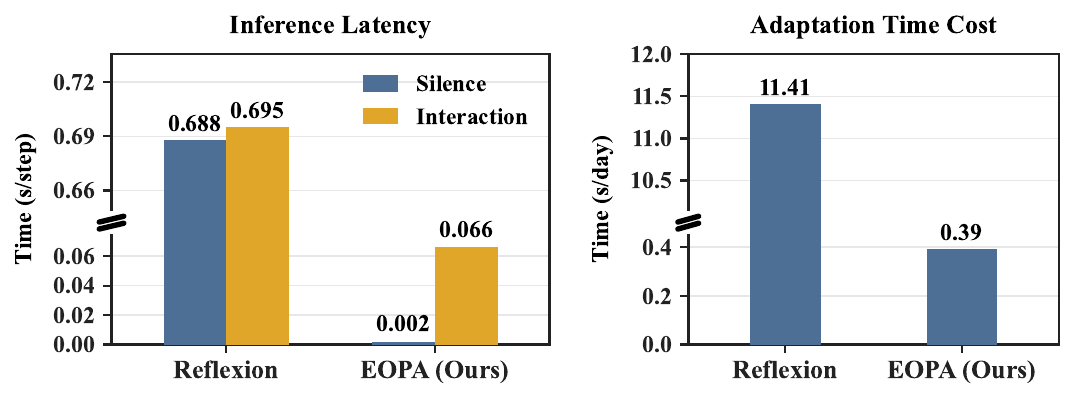}
    \caption{Inference latency and daily adaptation time cost.}
    \vspace{-10pt}
    \label{fig:efficiency_analysis}
\end{figure}

\begin{table}[t]
\centering
\scriptsize
\caption{Robustness of \methodname~under different feedback retention rates (random seed: 42).}
\label{tab:feedback_retention}
\setlength{\tabcolsep}{2.5pt}
\begin{tabular}{c|cccccc}
\toprule
Feedback Retained 
& F1 $\uparrow$
& F1@1 $\uparrow$
& F1@2 $\uparrow$
& Recall $\uparrow$
& Precision  $\uparrow$
& Accuracy $\uparrow$ \\
\midrule
100\% & 30.46 & 32.77 & 33.54 & 41.10 & 24.89 & 97.73\\
75\% & 27.30 & 29.28 & 30.12 & 35.32 & 23.33 & 97.73\\
50\% & 25.56 & 27.86 & 28.97 & 29.16 & 24.27 & 98.03\\
25\% & 13.35 & 14.61 & 15.49 & 12.50 & 16.83 & 98.40\\
\bottomrule
\end{tabular}
\vspace{-5pt}
\end{table}

\begin{table}[t]
\centering
\scriptsize
\caption{Discriminability of contextual evidence during online adaptation.}
\label{tab:discriminability}
\setlength{\tabcolsep}{11pt}
\begin{tabular}{cccc}
\toprule
Stage
& Evidence Gap $\uparrow$
& Pos. Above $\uparrow$
& Neg. Above $\downarrow$ \\
\midrule
Days 1--3  & 0.405 & 5.55\%  & 0.50\% \\
Days 4--7  & 1.365 & 46.00\% & 1.81\% \\
Days 8--14 & 2.097 & 53.31\% & 2.00\% \\
\bottomrule
\end{tabular}
\end{table}

\begin{table}[t]
\centering
\scriptsize
\caption{Comparison of Similar-State Decision Accuracy.}
\label{tab:similar_state_decision}
\resizebox{\columnwidth}{!}{
\begin{tabular}{ccccc}
\toprule
Method & Time Interval $\uparrow$ & Progress Stage $\uparrow$ & Activity Description $\uparrow$ & Avg. $\uparrow$ \\
\midrule
ProPerAssistant & 12.67 & 10.83 & 10.19 & 11.23 \\
EvoTest          & 20.88 & 13.47 & 16.78 & 17.04 \\
SCOPE            & 12.72 &  9.62 & 13.39 & 11.91 \\
Reflexion        & 24.31 & 18.44 & 25.42 & 22.72 \\
\textbf{EOPA (Ours)}
                 & \textbf{37.86}
                 & \textbf{35.61}
                 & \textbf{42.33}
                 & \textbf{38.60} \\
\bottomrule
\end{tabular}}
\vspace{-10pt}
\end{table}

\textbf{Inference and Adaptation Efficiency.}
Fig.~\ref{fig:efficiency_analysis} compares \methodname~with Reflexion in inference latency and daily adaptation time on identical hardware (Appendix~I). Latency is measured from receiving the user state to producing a silence decision or the first response chunk, after which downstream modules such as text-to-speech can begin streaming the response.
\methodname~reduces silence-step average latency
from 688 ms to 1.96 ms and interaction-step average
latency from 695 ms to 65.9 ms. This gain comes from decoupling timing decisions from response generation: the silence decision requires no LLM call, while interaction uses a compact prompt with a few demonstrations. In contrast, Reflexion invokes the LLM at every step with a longer prompt containing trajectory history and stored self-reflections.
\methodname~also reduces daily adaptation time from 11.41 to 0.39 s through lightweight evidence updates instead of LLM-based self-reflection.

\textbf{Robustness to Partial Feedback.}
In practice, users may not provide explicit feedback after every decision. We therefore use all feedback as the full-feedback setting and randomly retain 75\%, 50\%, or 25\% of the feedback for online adaptation. As in Table~\ref{tab:feedback_retention}, with 75\% and 50\% feedback, \methodname~achieves 27.30 and 25.56 F1, retaining 89.6\% and 83.9\% of its full-feedback performance. The decline mainly affects Recall. Precision remains stable, indicating more conservative decisions rather than increased unnecessary interruptions. Even with 25\% feedback, \methodname~achieves 13.35 F1, still outperforming Reflexion with full feedback (10.66). Overall, \methodname~remains robust to moderate feedback sparsity.

\textbf{Discriminability of Contextual Evidence.}
We evaluate whether the fused evidence score separates interaction-appropriate from silence-preferred states across the early (Days 1--3), middle (Days 4--7), and late (Days 8--14) stages. We report three metrics: Evidence Gap, which measures how much higher the average fused score for interaction-appropriate states is than that for silence-preferred states; Pos. Above, the proportion of interaction-appropriate states that exceed the decision threshold; and Neg. Above, the proportion of silence-preferred states that incorrectly exceed the threshold. As shown in Table~\ref{tab:discriminability}, Evidence Gap increases from 0.405 to 2.097 and Pos. Above from 5.55\% to 53.31\%, while Neg. Above remains low at 2.00\% in the late stage. These results show that accumulated feedback improves evidence separation and identifies more appropriate interaction moments while keeping false activations low.

\textbf{Fine-Grained Discrimination Among Similar States.}
To evaluate fine-grained discrimination, we extract state pairs that share the same temporal interval, activity-progress stage, or activity description but have different timing labels.
For each similarity criterion, we define the Similar-State Decision Accuracy (SSDA) as the proportion of state pairs for which a method makes the correct timing decision for both states. We average the three criterion-specific SSDA scores to obtain the average SSDA.
As in Table~\ref{tab:similar_state_decision}, \methodname~consistently outperforms all baselines and improves the average SSDA from 22.72\% for the Reflexion baseline to 38.60\%, demonstrating stronger discrimination among similar states with different interaction-timing needs.

\section{Conclusion}
We study feedback-driven online adaptation for personalized interaction initiation in proactive assistants. We propose \methodname, which organizes user feedback into temporal preference anchors and evidence-bearing activity prototypes to preserve fine-grained distinctions in interaction needs among similar user states. It derives temporal and activity evidence via user-prior-smoothed evidence estimation and uncertainty-guided scaling, and adaptively fuses them for interaction decisions. It then updates the carriers and decision parameters from feedback without LLM-based reasoning or retraining. Experiments on a ProPerSim-based benchmark show that \methodname~improves interaction-timing performance while reducing inference latency and online adaptation time.

\bibliography{reference}

@article{tu2025towards,
  title={Towards conversational diagnostic artificial intelligence},
  author={Tu, Tao and Schaekermann, Mike and Palepu, Anil and Saab, Khaled and Freyberg, Jan and Tanno, Ryutaro and Wang, Amy and Li, Brenna and Amin, Mohamed and Cheng, Yong and others},
  journal={Nature},
  volume={642},
  number={8067},
  pages={442--450},
  year={2025},
  publisher={Nature Publishing Group UK London}
}

@article{habicht2024closing,
  title={Closing the accessibility gap to mental health treatment with a personalized self-referral chatbot},
  author={Habicht, Johanna and Viswanathan, Sruthi and Carrington, Ben and Hauser, Tobias U and Harper, Ross and Rollwage, Max},
  journal={Nature medicine},
  volume={30},
  number={2},
  pages={595--602},
  year={2024},
  publisher={Nature Publishing Group US New York}
}

@article{algamdi2026behaviour,
  title={A behaviour-adaptive AI assistant enhancing accessibility and usability for blind users through real-time interaction personalization},
  author={Algamdi, Shabbab Ali},
  journal={Scientific Reports},
  year={2026},
  publisher={Nature Publishing Group UK London}
}

@inproceedings{shrivastava2025natural,
  title={Natural language processing for conversational AI: Chatbots and virtual assistants},
  author={Shrivastava, Neeraj and Tewari, Pushpa and Sujatha, S and Bogireddy, Srinivasa Rao and Varshney, Neeraj and Sharma, Vinod},
  booktitle={2025 IEEE International Conference on Interdisciplinary Approaches in Technology and Management for Social Innovation (IATMSI)},
  volume={3},
  pages={1--6},
  year={2025},
  organization={IEEE}
}

@inproceedings{yangcontextagent,
 author = {Yang, Bufang and Xu, Lilin and Zeng, Liekang and Liu, Kaiwei and Jiang, Siyang and Lu, Wenrui and Chen, Hongkai and Jiang, Xiaofan and Xing, Guoliang and Yan, Zhenyu},
 booktitle = {Advances in Neural Information Processing Systems},
 pages = {167509--167543},
 title = {ContextAgent: Context-Aware Proactive LLM Agents with Open-world Sensory Perceptions},
 volume = {38},
 year = {2025}
}

@inproceedings{luproactive,
  title={Proactive Agent: Shifting LLM Agents from Reactive Responses to Active Assistance},
  author={Lu, Yaxi and Yang, Shenzhi and Qian, Cheng and Chen, Guirong and Luo, Qinyu and Wu, Yesai and Wang, Huadong and Cong, Xin and Zhang, Zhong and Lin, Yankai and others},
  booktitle={The Thirteenth International Conference on Learning Representations},
  year={2025}
}

@inproceedings{
kim2026propersim,
title={ProPerSim: Developing Proactive and Personalized {AI} Assistants through User-Assistant Simulation},
author={Jiho Kim and Junseong Choi and Woosog Chay and Daeun Kyung and Yeonsu Kwon and Yohan Jo and Edward Choi},
booktitle={The Fourteenth International Conference on Learning Representations},
year={2026},
}

@inproceedings{zhao2024expel,
  title={Expel: Llm agents are experiential learners},
  author={Zhao, Andrew and Huang, Daniel and Xu, Quentin and Lin, Matthieu and Liu, Yong-Jin and Huang, Gao},
  booktitle={Proceedings of the AAAI Conference on Artificial Intelligence},
  volume={38},
  pages={19632--19642},
  year={2024}
}

@article{pei2025scope,
  title={Scope: Prompt evolution for enhancing agent effectiveness},
  author={Pei, Zehua and Zhen, Hui-Ling and Kai, Shixiong and Pan, Sinno Jialin and Wang, Yunhe and Yuan, Mingxuan and Yu, Bei},
  journal={arXiv preprint arXiv:2512.15374},
  year={2025}
}

@article{jeffreys1946invariant,
  title={An invariant form for the prior probability in estimation problems},
  author={Jeffreys, Harold},
  journal={Proceedings of the Royal Society of London. Series A. Mathematical and Physical Sciences},
  volume={186},
  number={1007},
  pages={453--461},
  year={1946},
  publisher={The Royal Society London}
}

@misc{qwen3.5,
    title  = {{Qwen3.5}: Towards Native Multimodal Agents},
    author = {{Qwen Team}},
    month  = {February},
    year   = {2026},
    url    = {https://qwen.ai/blog?id=qwen3.5}
}

@inproceedings{park2023generative,
  title={Generative agents: Interactive simulacra of human behavior},
  author={Park, Joon Sung and O'Brien, Joseph and Cai, Carrie Jun and Morris, Meredith Ringel and Liang, Percy and Bernstein, Michael S},
  booktitle={Proceedings of the 36th annual acm symposium on user interface software and technology},
  pages={1--22},
  year={2023}
}

@inproceedings{
he2025evotest,
title={EvoTest: Evolutionary Test-Time Learning for Self-Improving Agentic Systems},
author={Yufei He and Juncheng Liu and Yue Liu and Yibo Li and Tri Cao and Zhiyuan Hu and Xinxing Xu and Bryan Hooi},
booktitle={The Fourteenth International Conference on Learning Representations},
year={2026},
}

@misc{all_mpnet_base_v2,
    title = {{sentence-transformers/all-mpnet-base-v2}},
    author = {{Sentence Transformers}},
    howpublished = {\url{https://huggingface.co/sentence-transformers/all-mpnet-base-v2}},
    note = {Accessed: 2026-06-22},
    year = {2021}
}

@article{shinn2023reflexion,
  title={Reflexion: Language agents with verbal reinforcement learning},
  author={Shinn, Noah and Cassano, Federico and Gopinath, Ashwin and Narasimhan, Karthik and Yao, Shunyu},
  journal={Advances in neural information processing systems},
  volume={36},
  pages={8634--8652},
  year={2023}
}

@article{yang2025proagent,
  title={ProAgent: Harnessing On-Demand Sensory Contexts for Proactive LLM Agent Systems},
  author={Yang, Bufang and Xu, Lilin and Zeng, Liekang and Guo, Yunqi and Jiang, Siyang and Lu, Wenrui and Liu, Kaiwei and Xiang, Hancheng and Jiang, Xiaofan and Xing, Guoliang and others},
  journal={arXiv preprint arXiv:2512.06721},
  year={2025}
}

@article{he2026evoclinician,
  title={EvoClinician: A Self-Evolving Agent for Multi-Turn Medical Diagnosis via Test-Time Evolutionary Learning},
  author={He, Yufei and Liu, Juncheng and Hu, Zhiyuan and Chen, Yulin and Liu, Yue and Sui, Yuan and Li, Yibo and Chen, Nuo and Hu, Jun and Hooi, Bryan and others},
  journal={arXiv preprint arXiv:2601.22964},
  year={2026}
}

@article{pitis2024improving,
  title={Improving context-aware preference modeling for language models},
  author={Pitis, Silviu and Xiao, Ziang and Le Roux, Nicolas and Sordoni, Alessandro},
  journal={Advances in Neural Information Processing Systems},
  volume={37},
  pages={70793--70827},
  year={2024}
}

@article{chiang2022adaptive,
  title={An adaptive, context-aware, and stacked attention network-based recommendation system to capture users’ temporal preference},
  author={Chiang, Jung-Hsien and Ma, Chung-Yao and Wang, Chi-Shiang and Hao, Pei-Yi},
  journal={IEEE Transactions on Knowledge and Data Engineering},
  volume={35},
  number={4},
  pages={3404--3418},
  year={2022},
  publisher={IEEE}
}

@article{yu2023continuous,
  title={Continuous-time user preference modelling for temporal sets prediction},
  author={Yu, Le and Liu, Zihang and Sun, Leilei and Du, Bowen and Liu, Chuanren and Lv, Weifeng},
  journal={IEEE Transactions on Knowledge and Data Engineering},
  volume={36},
  number={4},
  pages={1475--1488},
  year={2023},
  publisher={IEEE}
}

@inproceedings{lin2024rella,
  title={Rella: Retrieval-enhanced large language models for lifelong sequential behavior comprehension in recommendation},
  author={Lin, Jianghao and Shan, Rong and Zhu, Chenxu and Du, Kounianhua and Chen, Bo and Quan, Shigang and Tang, Ruiming and Yu, Yong and Zhang, Weinan},
  booktitle={Proceedings of the ACM Web Conference 2024},
  pages={3497--3508},
  year={2024}
}

@article{he2009learning,
  title={Learning from imbalanced data},
  author={He, Haibo and Garcia, Edwardo A},
  journal={IEEE Transactions on knowledge and data engineering},
  volume={21},
  number={9},
  pages={1263--1284},
  year={2009},
  publisher={Ieee}
}

@book{gelman2013bayesian,
  title={Bayesian Data Analysis},
  author={Gelman, Andrew and Carlin, John B and Stern, Hal S and Dunson, David B and Vehtari, Aki and Rubin, Donald B},
  year={2013},
  publisher={CRC Press}
}

@inproceedings{wei2023online,
  title={Online prototype learning for online continual learning},
  author={Wei, Yujie and Ye, Jiaxin and Huang, Zhizhong and Zhang, Junping and Shan, Hongming},
  booktitle={Proceedings of the IEEE/CVF international conference on computer vision},
  pages={18764--18774},
  year={2023}
}

@inproceedings{dey2000cybreminder,
  title={Cybreminder: A context-aware system for supporting reminders},
  author={Dey, Anind K and Abowd, Gregory D},
  booktitle={International Symposium on Handheld and Ubiquitous Computing},
  pages={172--186},
  year={2000},
  organization={Springer}
}

@inproceedings{siewiorek2003sensay,
  title={SenSay: A Context-Aware Mobile Phone.},
  author={Siewiorek, Daniel P and Smailagic, Asim and Furukawa, Junichi and Krause, Andreas and Moraveji, Neema and Reiger, Kathryn and Shaffer, Jeremy and Wong, Fei Lung},
  booktitle={ISWC},
  volume={3},
  pages={248},
  year={2003}
}

@inproceedings{ho2005using,
  title={Using context-aware computing to reduce the perceived burden of interruptions from mobile devices},
  author={Ho, Joyce and Intille, Stephen S},
  booktitle={Proceedings of the SIGCHI conference on Human factors in computing systems},
  pages={909--918},
  year={2005}
}

@article{dorneich2012considering,
  title={Considering etiquette in the design of an adaptive system},
  author={Dorneich, Michael C and Ververs, Patricia May and Mathan, Santosh and Whitlow, Stephen and Hayes, Caroline C},
  journal={Journal of Cognitive Engineering and Decision Making},
  volume={6},
  number={2},
  pages={243--265},
  year={2012},
  publisher={Sage Publications Sage CA: Los Angeles, CA}
}

@article{lee2024aligning,
  title={Aligning to thousands of preferences via system message generalization},
  author={Lee, Seongyun and Park, Sue Hyun and Kim, Seungone and Seo, Minjoon},
  journal={Advances in Neural Information Processing Systems},
  volume={37},
  pages={73783--73829},
  year={2024}
}

@article{poddar2024personalizing,
  title={Personalizing reinforcement learning from human feedback with variational preference learning},
  author={Poddar, Sriyash and Wan, Yanming and Ivison, Hamish and Gupta, Abhishek and Jaques, Natasha},
  journal={Advances in Neural Information Processing Systems},
  volume={37},
  pages={52516--52544},
  year={2024}
}

@article{gao2024aligning,
  title={Aligning llm agents by learning latent preference from user edits},
  author={Gao, Ge and Taymanov, Alexey and Salinas, Eduardo and Mineiro, Paul and Misra, Dipendra},
  journal={Advances in neural information processing systems},
  volume={37},
  pages={136873--136896},
  year={2024}
}

@article{zeng2026glm,
  title={Glm-5: from vibe coding to agentic engineering},
  author={Zeng, Aohan and Lv, Xin and Hou, Zhenyu and Du, Zhengxiao and Zheng, Qinkai and Chen, Bin and Yin, Da and Ge, Chendi and Huang, Chenghua and Xie, Chengxing and others},
  journal={arXiv preprint arXiv:2602.15763},
  year={2026}
}

@article{xu2026deepseek,
  title={Deepseek-v4: Towards highly efficient million-token context intelligence},
  author={Xu, Anyi and Lin, Bangcai and Xue, Bing and Wang, Bingxuan and Xu, Bingzheng and Wu, Bochao and Zhang, Bowei and Lin, Chaofan and Dong, Chen and Ling, Chenchen and others},
  journal={arXiv preprint arXiv:2606.19348},
  year={2026}
}

@article{schobel2024charting,
  title={Charting the evolution and future of conversational agents: A research agenda along five waves and new frontiers},
  author={Sch{\"o}bel, Sofia and Schmitt, Anuschka and Benner, Dennis and Saqr, Mohammed and Janson, Andreas and Leimeister, Jan Marco},
  journal={Information Systems Frontiers},
  volume={26},
  number={2},
  pages={729--754},
  year={2024},
  publisher={Springer}
}

@inproceedings{zhou2024streaming,
  title={Streaming dense video captioning},
  author={Zhou, Xingyi and Arnab, Anurag and Buch, Shyamal and Yan, Shen and Myers, Austin and Xiong, Xuehan and Nagrani, Arsha and Schmid, Cordelia},
  booktitle={Proceedings of the IEEE/CVF Conference on Computer Vision and Pattern Recognition},
  pages={18243--18252},
  year={2024}
}

@article{yang2025qwen3,
  title={Qwen3 technical report},
  author={Yang, An and Li, Anfeng and Yang, Baosong and Zhang, Beichen and Hui, Binyuan and Zheng, Bo and Yu, Bowen and Gao, Chang and Huang, Chengen and Lv, Chenxu and others},
  journal={arXiv preprint arXiv:2505.09388},
  year={2025}
}

@article{csigdigital,
title={A survey on multimodal real-time interactive digital humans},
author={Du, Ruiqi and Yang, Boai and Zhou, Fengbo and Qu, Wei and Li, Tao},
journal={Journal of Image and Graphics},
pages={1-27},
year={2026},
}

@article{li2026challenges,
  title={Challenges and trends in egocentric vision: A survey},
  author={Li, Xiang and Qiu, Heqian and Wang, Lanxiao and Zhang, Hanwen and Qi, Chenghao and Han, Linfeng and Xiong, Huiyu and Li, Hongliang},
  journal={Machine Intelligence Research},
  volume={23},
  number={1},
  pages={1--33},
  year={2026},
  publisher={Springer}
}

\newpage

\appendix

\end{document}